\documentclass[prb,twocolumn,showpacs,floats,eqsecnum,amsmath,amssymb]{revtex4}
\usepackage[dvips]{graphicx}
\begin{document}
\title{Shot noise of a ferromagnetic nanowire with a domain wall}
\author{Babak Abdollahipour, Malek Zareyan and Nazila Asaadi}
\affiliation{Institute for Advanced Studies in Basic Sciences
(IASBS), P.O. Box 45195-1159, Zanjan 45195, Iran}

\begin{abstract}
We study shot noise of the spin-polarized current in a diffusive
ferromagnetic nanowire which contains a ballistic domain wall. We
find that the existence of a short domain wall influences strongly
the shot noise for sufficiently high spin-polarization of the
wire. Compared to the situation of absence of the domain wall, the
shot noise can be reduced or enhanced depending on the length of
the domain wall and its relative conductance.
\end{abstract}
\maketitle

\section{Introduction}

Electronic transport through ferromagnetic domain walls (DWs), the
regions with rotating magnetization vectors which connect two
homogeneous domains with miss-oriented magnetizations, has been
recently a subject of extensive investigations, both
theoretically\cite{Tatara08} and
experimentally\cite{Kent01,Marrows05}. This growing interest is
stimulated by the fundamental new physics raising from the dynamics
of spin of electron in DW\cite{Maekawa02}, as well as by the
potential applications in nanoelectronics and spintronics
devices\cite{Parkin02}. Among the others, recent experiments on
magnetic nanostructures and nanowires have revealed that the
presence of a DW may result in a magnetoresistance as large as
several hundreds or even thousands of
percents\cite{Garcia99,Chopra02,Ruster03}.
\par
In the bulk metallic ferromagnets like Fe, Co and Ni the so called
Bloch walls are the favor magnetic configurations where the rotating
magnetization vector is in the plane of DW. Such a DW is rather
thick with a length of about several hundreds $nm$. However, in
ferromagnetic nanowires the so called ``N\'{e}el wall''s are more
favored due to the transverse confinement\cite{Ebels00}. In a
``N\'{e}el wall'' the magnetization vector rotates in the plane
perpendicular to the plane of DW and the thickness of the DW can be
of order of $10~nm$. In ferromagnetic nanostructures, like a narrow
constriction between to wider domains, even sharper DWs with lengths
of the atomic scale can appear \cite{Bruno99,Pietzsch00,aui03}. In
the two latter cases the length of DW is usually smaller than the
electron mean free path of scattering from the static disorders, and
thus the electron transport is ballistic. Several theoretical works
have been devoted to studying contribution of the ballistic DWs on
the resistance of the nanowires and magnetic
nanostructures\cite{Hoof99,Brataas99,Zhuravlev03,Dugaev05}. In a
very thick DW the spin of the electron propagating across the wall
follows the magnetization direction quasi-adiabatically. Then
scattering of the electron from DW is very small and contribution of
the DW in the resistance is
negligible\cite{Cabrera74,Levy97,Tatara97,Gorkom99}. While for a
narrow DW the dynamic of spin of electron through the wall is not
adiabatic and the presence of the DW causes to considerable
scattering of electron. Calculations in the ballistic regime show an
increase of the resistance due to the
DW\cite{Imamura00,Dugaev03,Gopar04,Fallon04}.
\par
In spite of several theoretical and experimental studies of the
contribution of a DW on the average current, to our best
knowledge, there have been no works devoted to the fluctuation of
spin-polarized current in DWs. Low temperature temporal
fluctuation of the electrical current through a mesoscopic
conducting structure, the so called shot noise, provides valuable
information about the charge transport process which are not
extractable from the mean
conductance\cite{Blanter00,Belzig04,Zareyan05,Zareyan005,Hatami06}.
The aim of the present work is to study the effect of
spin-dependent scattering of electrons in a ferromagnetic DW on
the shot noise.
\par
We consider a ferromagnetic nanowire consisting of a $180^{\circ}$
ballistic DW connected to two diffusive domains with the
magnetization vectors aligned antiparallel to each other. Employing
the two-band Stoner Hamiltonian and within the scattering formalism,
we calculate the spin-dependent transmission coefficients of DW. The
resulting shot noise shows strong dependence on the size of the DW
as well as the degree of the spin-polarization of the nanowire. For
a thick DW where the spin-dynamic is dominated by the
quasi-adiabatic following of the local magnetization vector, the
shot noise has its normal value (shot noise of the wire without DW)
determined solely by the conductances of the diffusive domains and
the DW itself. However at lower thickness of DW, the shot noise
deviates significantly from the normal value depending on the
spin-polarization and the ratio of the conductance of the DW to the
domains. The interplay between diffusive transport at domains and
the noncollinear magnetization of the DW causes to reduction of the
shot noise below the normal value with varying the DW thickness.
\par
In the next section, we introduce a circuit which models the
diffusive ferromagnetic nanowire with a ballistic N\'{e}el DW. We
calculate the spin-polarized scattering coefficients of DW, which
are essential for the calculations of the contribution of DW in the
shot noise. Section III is devoted to developing formulas for the
average current and the shot noise of the nanowire. We analyze the
obtained results in section IV, for a full range of the DW
thickness, the spin-polarization and the relative conductance of DW.
Finally, in section V we give a conclusion.

%%%%%%%%%%%%%%%%%%%%%%%%%%%%%%%%%%%%%%%%%%%%%%%%%%%%%%%%%%%%%%%%%%%%%

\section{Modeling and the basic equations}

%%%%%%%%%%%%%%%%%%%%%%%%%%%%%%%%%%%%%%%%%%%%%%%%%%%%%%%%%%%%%%%%%%%%%

\begin{figure}
\centerline{\includegraphics[width=8cm,angle=0]{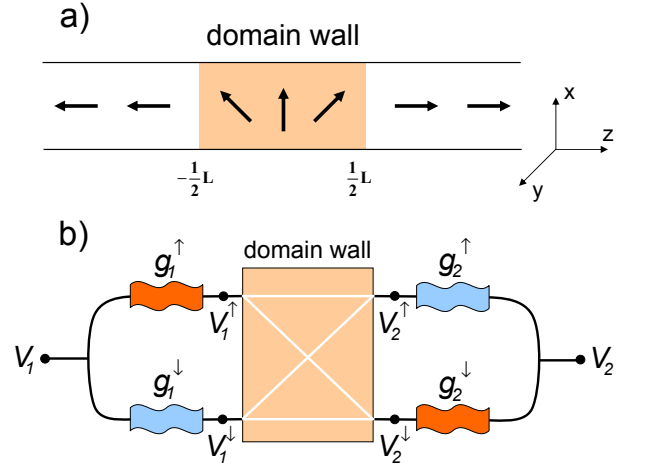}}
\caption{a) Schematic of a nanowire containing a domain wall and, b)
the corresponding circuite model of the nanowire.} \label{fig1}
\end{figure}

We consider a ferromagnetic nanowire consisting of two diffusive
ferromagnetic domains with antiparallel magnetization vectors which
are connected through a $180^{\circ}$ ballistic DW with length $L$.
Fig. 1. shows a sketch of the nanowire and the corresponding
circuite consisting of the two diffusive domains and the ballistic
DW. In the absence of extrinsic spin-flip scattering processes (for
instance due to magnetic impurities), each diffusive domain is
represented by two parallel spin-dependent conductances. In the
circuit model the DW is represented as a coherent four-terminal
scattering region which is connected through ideal ferromagnetic
leads to conducting elements of the two domains (see Fig.
\ref{fig1}b). We use the scattering approach and follow the
Refs.[\onlinecite{Levy97},\onlinecite{Gorkom99},\onlinecite{Gopar04}]
to calculate the scattering coefficients of the electron through the
DW. Within the two-band Stoner model, where the $d$-band electrons
are responsible for the magnetization and the current is carried by
the $s$-band electrons, we may describe transport of the electrons
through DW by an effective Hamiltonian of the form
\begin{equation}\label{eq:H}
\hat{H}=-\frac{\hbar^2}{2m}\nabla^2 +\frac{h_0}{2}
\hat{m}(\mathbf{r})\cdot\hat{\mbox{\boldmath $\sigma$}}\ ,
\end{equation}
where $h_0$ is the spin splitting of the $s$ electrons due to the
exchange coupling with the $d$ electrons and $\hat{\mbox{\boldmath
$\sigma$}}$ is the vector of the Pauli matrices.  First term of the
Hamiltonian is the kinetic energy of electron and the second part
represents the interaction of the spin of electron with the local
magnetization oriented in the direction of the unit vector
$\hat{m}(\mathbf{r})$. In this work we consider a ``N\'{e}el wall'',
which is more common configuration in laterally confined
ferromagnetic nanowires. We assume that
$\hat{m}(\mathbf{r})=\left(m_{x}(z), 0, m_{z}(z)\right)$ varies
along the $z$ direction (wire axis). In the left and right domains
the magnetization vectors are aligned along the $-z$ and $z$ axis,
respectively. The assumption $\hat{m}(\mathbf{r})=\hat{m}(z)$ allows
us to separate the transverse and longitudinal parts of the
Hamiltonian (\ref{eq:H}). In the transverse direction the motion of
the electron is quantized with an energy denoted by $E_{\bot}$. From
Eq. (\ref{eq:H}) the effective Schrodinger equation for the
longitudinal motion of electron is obtained as
\begin{equation}\label{scequation}
\left(-\frac{\hbar^2}{2m}\frac{d^{2}}{d z^{2}}+\frac{h_0}{2}
\hat{m}(z)\cdot\hat{\mbox{\boldmath
$\sigma$}}\right)\mathbf{\Psi}(z) =\varepsilon\mathbf{\Psi}(z) \ ,
\end{equation}
where $\varepsilon=E-E_{\bot}$ is the longitudinal energy. To be
specific we consider a trigonometric magnetization profile in the
DW, which is defined by \cite{Gopar04}
\begin{equation}\label{trigonometric-wall}
\hat{m}(z)=\left\{
\begin{array}{cl}
\left(\cos\frac{\pi z}{L},\;0,\; \sin\frac{\pi z}{L}\right),&
{\rm for}\; |z|\le L/2 \\
\left(\;0 \; ,\;\; 0\;\; , \;{\rm sgn}(z)\; \right), &{\rm for}\;
|z| > L/2
\end{array}\right.
\end{equation}
The advantage of this choice is that, it admits an exact solution
for the wave functions inside the DW.
\par
In order to obtain the spin-dependent transmission and reflection
coefficients we have to solve the Schrodinger equation
(\ref{scequation}) in different regions and then match the solutions
of different regions at the boundaries ($z=\pm L/2$). In the domains
$z<-L/2$ and $z>L/2$ the eigenfunctions have the form $\Psi=
e^{ik_{\sigma}z}|\hat{z},\sigma\rangle$, where $\sigma=\pm 1$ denote
up and down spin directions and $|\hat{z},\sigma\rangle$ are the
spin states when spin quantization axis is chosen to be the $z$
axis. The longitudinal wave vector for spin-$\sigma$ electrons is
given by
$k_{\sigma}=\sqrt{\frac{2m}{\hbar^{2}}(\varepsilon+\sigma\frac{h_0}{2})}$
. To find the eigenfunctions in the DW we do a transformation in
spin space from the fixed reference frame to the rotated frame,
which is in the direction of the local magnetization vector
$\hat{m}(z)$. In our representation it is given by a rotation about
the $y$ axis $\hat{R}=exp\left(-i\sigma_{y}\theta/2\right)$, where
$\theta=\tan^{-1}(m_{x}/m_{z})=\frac{\pi}{2}-\frac{\pi z}{L}$ is the
angle of the magnetization vector respect to the $z$ axis at the
point $z$. The Hamiltonian in the rotating frame
$\hat{\mathcal{H}}_{r}=\hat{R}^{-1}\hat{H}_{z}\hat{R}$, where
$\hat{H}_{z}$ is the longitudinal part of the Hamiltonian at fixed
frame, takes the form
\begin{equation}\label{rotatingh}
\hat{\mathcal{H}}_{r}=-\frac{\hbar^2}{2m}\left(\frac{d^2}{dz^2}-\omega^2\right)
+i\omega\sigma_{y}\frac{\hbar^2}{m}\frac{d}{dz}+\frac{h_0}{2}\sigma_{z}\
,
\end{equation}
which does not depend on $z$ explicitly. Here
$\omega=\frac{1}{2}\frac{d}{dz}\theta=-\frac{\pi}{2L}$. The
eigenfunctions of this Hamiltonian have the forms
\begin{eqnarray}\label{wavefunction}
\Phi_{\sigma}(z)= \left(
\begin{array}{cc}
\begin{array}{c}
\tilde{u}_{\sigma}
\end{array}
\\
\begin{array}{c}
\\
\tilde{v}_{\sigma}
\end{array}
\end{array}
\right)e^{iq_{\sigma}z},
\end{eqnarray}
where
\begin{eqnarray}&&
\tilde{u}^{2}_{\sigma}=\frac{1}{2}\left[1-
\frac{mh_0/\hbar^2}{2\omega^2+\sigma\sqrt{\frac{8m\varepsilon\omega^2}{\hbar^2}
+\left(\frac{mh_0}{\hbar^2}\right)^2}}\right]\ , \nonumber\\&&
\tilde{v}^{2}_{\sigma}=\frac{1}{2}\left[1+
\frac{mh_0/\hbar^2}{2\omega^2+\sigma\sqrt{\frac{8m\varepsilon\omega^2}{\hbar^2}
+\left(\frac{mh_0}{\hbar^2}\right)^2}}\right]\ ,
\end{eqnarray}
and the longitudinal wave vectors are
\begin{equation}\label{wavevector}
q_{\sigma}=\left[\left(\frac{2m\varepsilon}{\hbar^{2}}+
\omega^{2}\right)+\sigma\sqrt{\frac{8m\varepsilon\omega^2}{\hbar^2}
+\left(\frac{mh_0}{\hbar^2}\right)^2}~\right]^{1/2}.
\end{equation}
The wave functions in the fixed reference frame (along z axis) are
obtained from the relations
$\Psi_{\sigma}(z)=\hat{R}\Phi_{\sigma}(z)$. Now, if we consider a
spin up electron incident to the DW from the left domain, the wave
functions in three regions have the following forms
\begin{equation}
\Psi_1=e^{ik_{+}z}\left(\begin{array}{cl} 1 \\ 0
\end{array}\right) +
r_{\uparrow\uparrow}e^{-ik_{+}z}\left(\begin{array}{cl} 1 \\ 0
\end{array}\right)+ r_{\downarrow\uparrow}e^{-ik_{-}z}\left(
\begin{array}{cl}
0 \\ 1 \end{array}\right)\ ,
\end{equation}
for $z<-L/2$,
\begin{eqnarray}
\Psi_2&=&c_{1}e^{iq_{+}z}\left(\begin{array}{cl} u_{+} \\
v_{+} \end{array}\right) +
c_{2}e^{-iq_{+}z}\left(\begin{array}{cl} u_{+} \\
v_{+} \end{array}\right)
\nonumber\\&+&
c_{3}e^{iq_{-}z}\left(
\begin{array}{cl} u_{-} \\v_{-} \end{array}\right)
+ c_{4}e^{-iq_{-}z}\left(
\begin{array}{cl}u_{-} \\ v_{-}\end{array}\right)\ ,
\end{eqnarray}
for $-L/2\leq z\leq L/2$ and
\begin{equation}
\Psi_3=t_{\uparrow\uparrow}e^{ik_{-}z}\left(\begin{array}{cl} 1 \\
0
\end{array}\right)+ t_{\downarrow\uparrow}e^{ik_{+}z}\left(
\begin{array}{cl}
0 \\ 1 \end{array}\right)\ ,
\end{equation}
for $z>L/2$. Here $t_{\uparrow\uparrow}$ ($r_{\uparrow\uparrow}$)
and $t_{\downarrow\uparrow}$ ($r_{\downarrow\uparrow}$) are the
spin-conserved and spin-flip transmission (reflection) coefficients,
respectively. Imposing the condition of the continuity of the wave
functions and their first derivatives at the boundaries ($z=\pm
L/2$), we obtain these scattering coefficients. The obtained
expressions are too lengthly to be given here. We only mention some
properties of the reflection and transmission probabilities. Both
spin-conserved $|r_{\uparrow\uparrow}|^{2}$ and spin-flip
$|r_{\downarrow\uparrow}|^{2}$ reflection probabilities are small
except for a narrow DW of the length $L\leq \ell_{DW}$, where
$\ell_{DW}=\frac{2}{k_F}(\frac{2E_F}{h_0})^{1/2}$ is the
spin-polarization dependent length scale. The spin-conserved
transmission probability $|t_{\uparrow\uparrow}|^{2}$, is close to
unity for a narrow DW, and has a diminishing behavior with
increasing $L/ \ell_{DW}$ toward a vanishing value for a thick DW.
In contrast the spin-flip transmission probability
$|t_{\downarrow\uparrow}|^{2}$, has an appreciable value for a
sizable DW of $L\geq \ell_{DW}$, where the spin of electron has
enough time to follow adiabatically the local magnetization
direction. In the limit of $L\gg \ell_{DW}$ the electron transport
is mainly realized through two independent channels connecting the
majority and minority spin electrons in two domains.

%%%%%%%%%%%%%%%%%%%%%%%%%%%%%%%%%%%%%%%%%%%%%%%%%%%%%%%%%%%%%%%%%%%%%

\section{Average current and Shot noise}\label{sec:circuit model}

%%%%%%%%%%%%%%%%%%%%%%%%%%%%%%%%%%%%%%%%%%%%%%%%%%%%%%%%%%%%%%%%%%%%%

To express the average current and shot noise of the nanowire in
terms of the scattering coefficients derived in the previous
section, we use the the circuit model of nanowire shown in Fig.
\ref{fig1}b. In the circuite the diffusive domains are modeled by
two parallel conductances for up and down spin electrons, denoted by
$g_{\alpha}^{\sigma}$, where $\alpha=1,2$ labels the left and right
domains, respectively. These conductances have connected to the left
and right reservoirs with fixed voltages $V_1$ and $V_2$. The DW has
been shown as a four terminal device connected via the nodes
($\alpha$,$\sigma$) to the same domains through the ideal
ferromagnetic leads. For simplicity we consider a symmetric
structure for that
$g_{1}^{\uparrow(\downarrow)}=g_{2}^{\downarrow(\uparrow)}=g_{+(-)}$.
Noting the fact that the difference of the Fermi level density of
states for the majority and minority spin electrons in the domains
is proportional to the ratio $h_0/2E_F$, the spin-dependent
conductances are approximated by
\begin{equation}\label{resistances}
g_{\pm}=\frac{g_{F}}{2}(1\pm\frac{h_0}{2E_F})\ ,
\end{equation}
where $g_F=g_{+}+g_{-}$ is the total conductance of each diffusive
ferromagnetic domain. The conductance of the ballistic leads is
given by  $g_0=N e^2/h$, where the total number of open channels in
the leads $N=N_{+}+N_{-}$ is the sum of the number of open channels
for two spin states electrons $N_{\sigma}$. The value of $g_0/g_F$
is normally large in ferromagnetic domain structures.
\par
We derive the average current and the shot noise power from  the
spin-dependent Landauer-Buttiker formula \cite{Buttiker92} and the
scattering coefficients obtained in the previous section. The
current operator for spin-$\sigma$ electrons flowing through
terminal $\alpha$ is defined as
\begin{eqnarray}\label{currentoperator}
\hat{i}_{\alpha}^{\sigma}(t)&=&\frac{e}{h} \sum_{n=1}^{N_{\sigma}}
\int \!\! \int d\varepsilon\,d\varepsilon' \,
e^{i(\varepsilon-\varepsilon')t/\hbar} \nonumber\\&&
\left[\hat{a}_{\alpha n}^{\sigma
\dagger}(\varepsilon)\hat{a}_{\alpha n}^{\sigma}(\varepsilon')
-\hat{b}_{\alpha n}^{\sigma \dagger} (\varepsilon) \hat{b}_{\alpha
n}^{\sigma}(\varepsilon')\right]\ ,
\end{eqnarray}
where the operator ${\hat{a}^{\sigma^{\dagger}}}_{\alpha
n}(\varepsilon)$ $({\hat{a}^{\sigma}}_{\alpha n}(\varepsilon))$
creates (annihilates) outgoing electron from terminal $\alpha$ in
the n{\it th} channel with energy $\varepsilon$. Similarly,
${\hat{b}^{\sigma^{\dagger}}}_{\alpha n}$,
(${\hat{b}^{\sigma}}_{\alpha n}$) denotes creation (annihilates)
operator for a spin-$\sigma$ incoming electron in the terminal
$\alpha$. The corresponding average current reads
\begin{equation}\label{current}
i^{\sigma}_{\alpha}=\sum_{\rho,\beta}
G^{\sigma\rho}_{\alpha\beta}V^{\rho}_{\beta}\ ,
\end{equation}
where $\alpha$, $\beta$ stand for domains and $\sigma$, $\rho$ for
spin directions; $V^{\rho}_{\beta}$ are spin-$\rho$ voltages at the
connecting nodes between the domain $\beta$ and the DW, and
$G^{\sigma\rho}_{\alpha\beta}$ are elements of the conductance
matrix defined by
\begin{equation}\label{conductivity}
G^{\sigma\rho}_{\alpha\beta}=\frac{e^{2}}{h} Tr\left[\delta_{\alpha
\beta}\delta_{\sigma\rho}-\left(s^{\sigma\rho}_{\alpha
\beta}\right)^{\dagger}s^{\sigma\rho}_{\alpha \beta}\right]\ ,
\end{equation}
where $s^{\sigma\rho}_{\alpha \beta}$ are elements of the scattering
matrix of the DW at the Fermi energy. Considering that the number of
channels is very large and using the fact that the density of states
for a 2D system is constant we can change the $Tr$ to integral over
$\varepsilon$ in calculating the traces. Then we can write
\begin{equation}
Tr[S^{\sigma\rho^{\dagger}}_{\alpha \beta}S^{\sigma\rho}_{\alpha
\beta}]=\frac{N}{2E_F}\int^{E_F}_{-h_0/2} d\varepsilon
\left[S^{\sigma\rho^{\dagger}}_{\alpha
\beta}(\varepsilon)S^{\sigma\rho}_{\alpha
\beta}(\varepsilon)\right]\ .
\end{equation}
The corresponding expression for the zero-frequency correlation of
current fluctuations $S^{\sigma\sigma^{\prime}}_{\alpha
\alpha^{\prime}}=2\int dt \langle\delta i^{\sigma}_{\alpha}(t)\delta
i^{\sigma^{\prime}}_{\alpha^{\prime}}(0)\rangle$ is expressed as
\begin{eqnarray}\label{correlations-domain}&&
S^{\sigma\sigma^{\prime}}_{\alpha \alpha^{\prime}}=\frac{2
e^{2}}{h}\sum_{\gamma, \gamma^{\prime}}\sum_{\rho, \rho^{\prime}}
\nonumber\\&& Tr\left[\left(s^{\sigma\rho}_{\alpha
\gamma}\right)^{\dagger}s^{\sigma\rho^{\prime}}_{\alpha
\gamma^{\prime}}\left(s^{\sigma^{\prime}\rho^{\prime}}_{\alpha^{\prime}
\gamma^{\prime}}\right)^{\dagger}s^{\sigma^{\prime}\rho^{\prime}}_{\alpha^{\prime}
\gamma}\right]\left| V^{\rho}_{\gamma}-
V^{\rho^{\prime}}_{\gamma^{\prime}}\right|\ .
\end{eqnarray}
On the other hand the average current for spin-$\sigma$ electrons
through the domain $\alpha$ can be obtained via the relation
\begin{equation}\label{volages}
I_{\alpha}^{\sigma}=g_{\alpha}^{\sigma}\left(V_{\alpha}-V_{\alpha}^{\sigma}\right).
\end{equation}
Applying the conservation rule for spin-$\sigma$ current flowing
into the nodes ($\alpha$,$\sigma$) and using Eqs. (\ref{current})
and (\ref{volages}) we find
\begin{equation}\label{average-current}
g_{\alpha}^{\sigma}V_{\alpha}^{\sigma}+
\sum_{\rho,\beta}G^{\sigma\rho}_{\alpha\beta}V^{\rho}_{\beta}=
g_{\alpha}^{\sigma}V_{\alpha}\ .
\end{equation}
The solution of this matrix equation give us $V_{\alpha}^{\sigma}$
in terms of the voltage difference ($V_1-V_2$), the conductances
$g_{\alpha}^{\sigma}$ and $G^{\sigma\rho}_{\alpha\beta}$, from which
and using Eqs. (\ref{current}, \ref{correlations-domain}) we can
calculate the spin-resolved average currents and the contribution of
DW to the correlations of current fluctuations in different nodes.
\par
To calculate the noise power of the total charge current in $\alpha$
reservoir
$I_{\alpha}=I_{\alpha}^{\uparrow}+I_{\alpha}^{\downarrow}$, we
should include the effect of the voltage fluctuations in the nodes
as well as the fluctuations due to the scattering inside the
domains. Using equation (\ref{volages}) and denoting the intrinsic
fluctuations of the current due to the scattering inside the domain
by $\delta I^{\sigma}_{\alpha}$, we can write the total fluctuations
of the spin-$\sigma$ current coming from $\alpha$ domain to the
($\alpha$,$\sigma$) node as
\begin{equation}\label{cfluctuation1}
\Delta I^{\sigma}_{\alpha}=-g^{\sigma}_{\alpha}\delta
V^{\sigma}_{\alpha}+\delta I^{\sigma}_{\alpha}\ .
\end{equation}
where $\delta V^{\sigma}_{\alpha}$ is the voltage fluctuation in the
node ($\alpha$,$\sigma$). We notice that the voltages of reservoirs
$V_{\alpha}$ are constant. At the same time from Eq. (\ref{current})
we obtain the total fluctuation of the current flowing through the
node ($\alpha$,$\sigma$) in terms of the current fluctuation due to
scattering from DW:
\begin{equation}\label{current-fluctuation-domain}
\Delta i^{\sigma}_{\alpha}=\sum_{\rho,\beta}
G^{\sigma\rho}_{\alpha\beta}\delta V^{\rho}_{\beta}+\delta
i^{\sigma}_{\alpha}\ .
\end{equation}
where $\delta i^{\sigma}_{\alpha}$ is the intrinsic current
fluctuation of DW. Applying the conservation rule for the temporal
fluctuations of the currents $\Delta I^{\sigma}_{\alpha}=\Delta
i^{\sigma}_{\alpha}$, we obtain
\begin{equation}\label{current-fluctuation}
g^{\sigma}_{\alpha}\delta V^{\sigma}_{\alpha}+\sum_{\rho,\beta}
G^{\sigma\rho}_{\alpha\beta}\delta V^{\rho}_{\beta}=\delta
I^{\sigma}_{\alpha}-\delta i^{\sigma}_{\alpha}\ .
\end{equation}
By solving the above matrix equation we find $\delta
V^{\sigma}_{\alpha}$ in terms of the $\delta I^{\sigma}_{\alpha}$
and $\delta i^{\sigma}_{\alpha}$. The shot noise of the total
current in $\alpha$ reservoir is given by
\begin{equation}\label{total-shot noise}
\mathbf{S}_{\alpha \alpha}=\mathbf{S}^{\uparrow\uparrow}_{\alpha
\alpha}+\mathbf{S}^{\uparrow\downarrow}_{\alpha
\alpha}+\mathbf{S}^{\downarrow\uparrow}_{\alpha
\alpha}+\mathbf{S}^{\downarrow\downarrow}_{\alpha \alpha}\ ,
\end{equation}
where $\mathbf{S}^{\sigma\sigma^{\prime}}_{\alpha \alpha}=2\int dt
\langle\Delta I^{\sigma}_{\alpha}(t)\Delta
I^{\sigma^{\prime}}_{\alpha}(0)\rangle$, and is expressed in terms
of the correlations of the current fluctuations $\delta
I^{\sigma}_{\alpha}$ and $\delta i^{\sigma}_{\alpha}$. The
correlations of the currents fluctuations $\delta
i^{\sigma}_{\alpha}$ are given by Eq. (\ref{correlations-domain}),
and for the diffusive domains we have the following result for the
correlations of the currents $\delta I^{\sigma}_{\alpha}$
\begin{equation}\label{current-corralation}
2\int dt\left\langle\delta I^{\sigma}_{\alpha}(t) \delta
I^{\sigma\prime}_{\alpha\prime}(0)\right\rangle=\frac{1}{3}
g^{\sigma}_{\alpha}\left|V_{\alpha}-V^{\sigma}_{\alpha}\right|
\delta_{\alpha\alpha\prime}\delta_{\sigma\sigma\prime}\ .
\end{equation}
Using Eqs. (\ref{volages}), (\ref{total-shot noise}),
(\ref{current-corralation}) and (\ref{correlations-domain}) we can
calculate the average current and the shot noise of the nanowire in
terms of the system parameters. In the next section we discuss the
results for average current and Fano factor defined by
$F=\mathbf{S}_{\alpha\alpha}/2eI_{\alpha}$.
\begin{figure}
\centerline{\includegraphics[width=8cm,angle=0]{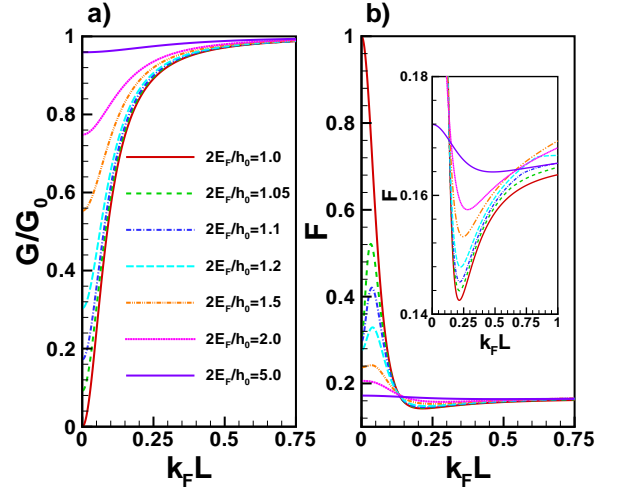}} \caption{
Conductance of the ferromagnetic nanowire with a domain wall
relative to its conductance without domain wall (a), and The Fano
factor of the nanowire (b), in terms of the domain wall length for
different values of $2E_F/h_0$ and $g_0/g_F=80$. Inset in the figure
(b) shows the minimum of the Fano factor.} \label{fig2}
\end{figure}
%

%%%%%%%%%%%%%%%%%%%%%%%%%%%%%%%%%%%%%%%%%%%%%%%%%%%%%%%%%%%%%%%%%%%%%%%

\section{Results and Discussions}\label{sec:results}

%%%%%%%%%%%%%%%%%%%%%%%%%%%%%%%%%%%%%%%%%%%%%%%%%%%%%%%%%%%%%%%%%%%%%%%

The average current and Fano factor of the nanowire can be expressed
in terms of the three dimensionless parameters $k_FL$, $2 E_{F}/h_0$
and $g_0/g_F$ which respectively characterize the thickness of the
DW, the degree of spin-polarization of the nanowire and the relative
conductance of the ballistic DW and the two domains.
\par
\begin{figure}
\centerline{\includegraphics[width=8cm,angle=0]{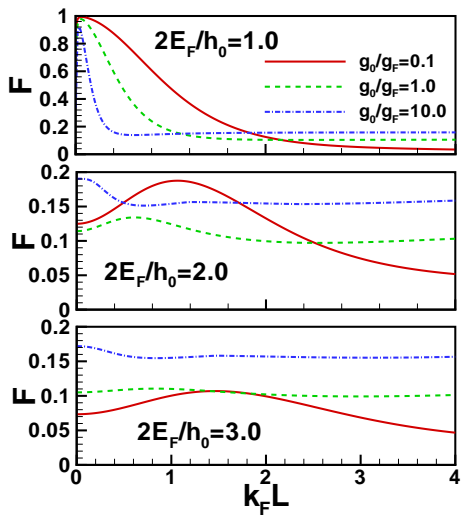}} \caption{
Fano factor of the ferromagnetic nanowire with a domain wall for
three values of $g_0/g_F$ and $2E_F/h_0$.} \label{fig3}
\end{figure}
Let us start with analyzing the effect of DW in the conductance of
the nanowire. In Fig. \ref{fig2}a, we have plotted the conductance
of the nanowire $G$ normalized to its conductance in the absence of
DW, $G_0$, versus $k_FL$ for $g_0/g_F=80$ and different values of
$2E_F/h_0$. Conductance of the nanowire rises with increasing the
thickness of the DW, which means that the presence of the DW always
increases resistance of the nanowire. Considerable change in the
conductance occurs for a short DW. The corresponding variation of
the Fano factor is shown in Fig. \ref{fig2}b. For the half metal
nanowire ($2E_F/h_0=1$) the Fano factor approaches its maximum
Poissonian value ($F=1$) at small lengths but decreases rapidly by
increasing $k_FL$. In a fully polarized nanowire at small lengths
the DW acts as a tunnel barrier and causes to Poissonian shot noise.
A similar effect is seen in the FNF spin valve
structures\cite{Zareyan05}. The Fano factor passes through a smooth
minimum before taking its normal value in the limit of $k_FL\gg 1$.
For smaller spin-polarizations (greater values of the $2E_F/h_0$)
the Fano factor shows a maximum smaller than one which occurs at a
finite $k_FL$. The smooth minima at a finite length of DW occurs for
smaller spin-polarizations as for the case of $2E_F/h_0=1$, as is
shown in the inset of Fig. \ref{fig2}b. This observation that the
Fano factor decreases below its normal value for specific lengths of
DW can be attributed to the noncollinear change of the magnetization
at DW. Increasing $k_FL$ causes a decrease in the spin conserving
transmission coefficient and an increase in the spin-flip
transmission coefficient. This tends to decrease the Fano factor. At
the same time by increasing the length of the DW, its conductance
increases and consequently the contribution of the domains in the
Fano factor increases. Competition of these two effects leads to
generation of the minimum in the Fano factor, such that $F$ goes
below the normal value. In this regime while the transport of
electron through DW is closely ballistic giving rise to a negligibly
small shot noise, its conductance has a sizable value to have a
significant contribution to the Fano factor of the whole structure.
A similar behavior has been seen in the noncollinear FNF systems
with diffusive junctions\cite{Tserkovnyak01,Abdollahi06}, where the
Fano factor reduces below its collinear value at specific values of
the relative angle of the magnetization vectors. In the limit of
large $k_FL$ the Fano factor tends to its normal value (single
domain ferromagnetic nanowire) given by
\begin{eqnarray}\label{long-limit}
F&=&\frac{\sum_{\sigma}F^{\sigma}I^{\sigma}}{\sum_{\sigma}I^{\sigma}}\
,
\nonumber\\
F^{\sigma}&=&\frac{(F_1^{\sigma}R_1^{\sigma
2}+F_{DW}^{\sigma}R_{DW}^{\sigma 2} +F_2^{\sigma}R_2^{\sigma
2})}{(R_1^{\sigma}+R_{DW}^{\sigma}+R_2^{\sigma})^2}\ ,
\end{eqnarray}
where $\sigma$ denotes majority and minority spin sub-bands,
$F_{1,2}^{\sigma}=1/3$ and $R_{1,2}^{\sigma}=1/g_{\sigma}$ are Fano
factors and resistances of the domains, $F_{DW}^{\sigma}=0$ and
$R_{DW}^{\sigma}=1/g_0$ are those of very thick DW. This expression
coincides with the results which is obtained by extending the
formula derived by Beenakker and Buttiker for Fano factor of a wire
consisting of a series of phase coherent segments in the inelastic
regime\cite{Beenakker92}. At this limit electrons pass adiabatically
trough the DW and mixing between majority and minority spin
sub-bands is negligible. Thus system behaves like a single domain
and DW has no considerable effect on the conduction.
\par
Let us now consider the effect of varying the relative conductances
of domains $g_0/g_F$ on the Fano factor. This is shown in Fig.
\ref{fig3} for different $2E_F/h_0$. The main feature is that by
decreasing $g_0/g_F$ the length scale over which $F$ has appreciable
variation increases. There are two contributions in the total shot
noise of the nanowire. One is the domains contribution which is
independent of the DW length and the other one is due to the DW and
depends on the length of the DW. The relative importance of them is
determined by voltage drops at these elements. For large $g_0/g_F$
domains act as resistive elements of the nanowire and cause to
lowering the voltage drop at DW and thus reducing the importance of
the DW contribution. As it is seen in the figure (\ref{fig3}) at
this limit the Fano factor shows small variations whit length. On
the opposite limit when $g_0/g_F$ is small the DW has dominant
contribution at the shot noise and the the Fano factor shows
considerable variations whit length of DW.

%%%%%%%%%%%%%%%%%%%%%%%%%%%%%%%%%%%%%%%%%%%%%%%%%%%%%%%%%%%%%%%%%%%%%

\section{Conclusion}

%%%%%%%%%%%%%%%%%%%%%%%%%%%%%%%%%%%%%%%%%%%%%%%%%%%%%%%%%%%%%%%%%%%%%

In conclusion, we have investigated the effect of a ballistic
domain wall on the spin-polarized shot noise of a ferromagnetic
nanowire. We have considered the inelastic regime where the
diffusive domains and the ballistic DW can be treated as the
separated coherent segments. Using the two-band Stoner model for a
N\'{e}el type trigonometric profile of the magnetization vector,
we have obtained that Fano factor changes significantly with
respect to its value in the absence of DW for a high
spin-polarization of the wire and when DW is short enough. A
remarkable result is that the presence of DW can cause both
reduction or enhancement of the shot noise, depending on its
length, the spin-polarization and the relative conductances of the
domains and DW. Here we considered an especial type of the profile
for the DW. The realistic profile of a DW can be different. Since
different profiles for the DW do not change the scattering
coefficients qualitatively\cite{Gopar04}, we expect that the
simplified profile we considered here will capture the essential
physics and the effect of considering other profiles to our
results for shot noise will be minor and quantitative.
\par
With the new developed techniques for measurement of the shot
noise in various systems especially the magnetic tunnel
junctions\cite{Guerrero08} and recent progresses in fabricating
and controlling different types of DWs\cite{Marrows05}, the shot
noise measurement in DW seems to be feasible. Such future
measurements can verify our results experimentally.

%%%%%%%%%%%%%%%%%%%%%%%%%%%%%%%%%%%%%%%%%%%%%%%%%%%%%%%%%%%%%%%%%%%%%%
%%%%%%%%%%%%%%%%%%%%%%%%%%%%%%%%%%%%%%%%%%%%%%%%%%%%%%%%%%%%%%%%%%%%%%

%\begin{acknowledgments}
%
%\end{acknowledgments}

\end{document}